\documentstyle[twoside,fleqn,espcrc2]{article}

\newcommand{\be}{\begin{equation}}
\newcommand{\ee}{\end{equation}}
\newcommand{\ba}{\begin{eqnarray}}
\newcommand{\ea}{\end{eqnarray}}
\newcommand{\AmS}{{\protect\the\textfont2
  A\kern-.1667em\lower.5ex\hbox{M}\kern-.125emS}}

\begin{document}
\hbox{\hskip 12cm ROM2F-97/4  \hfil}
\vskip 1.8cm
\begin{center}
{\Large  \bf  Surprises \ in \ Open-String \ Perturbation \ Theory}

\vspace{2cm}

{\large \large  Augusto Sagnotti}

\vspace{1.2cm}

{\sl Dipartimento di Fisica\\
Universit{\`a} di Roma \ ``Tor Vergata'' \\
I.N.F.N.\ - \ Sezione di Roma \ ``Tor Vergata'' \\
Via della Ricerca Scientifica, 1 \ \
00133 \ Roma \ \ ITALY}
\vspace{0.4cm}
\vskip 1.2cm
{\bf Abstract}
\end{center}
\vskip 18pt
{
The perturbative analysis of models of open and closed
superstrings presents a number of surprises.  For instance, variable
numbers of antisymmetric tensors ensure their consistency via
generalized Green-Schwarz cancellations and a novel type of
singularity occurs in their moduli spaces.  All these features are
related, in one way or another, to the presence of boundaries on the
world sheet or, equivalently, of extended objects (branes) interacting
with the bulk theory in space time.  String dualities have largely
widened the interest in these models, that exhibit a wealth of generic
non-perturbative features of String Theory.}
\vskip 1.8cm
\begin{center}
Contribution to the Proceedings of the XXX Ahrenshoop Meeting,
Buckow (Berlin), August 1996
\end{center}
\vfill\eject

\title{Surprises \ in \ Open-String \ Perturbation \ Theory}

\author{Augusto Sagnotti\address{Dipartimento di Fisica,  Universit\`a
di Roma ``Tor Vergata'' \\ 
        I.N.F.N., Sezione di Roma ``Tor Vergata''\\ Via della Ricerca
Scientifica, 1 \ \ 00133  \ Roma \ ITALY}%
        \thanks{Work supported in part by E.E.C. Grant
CHRX-CT93-0340.}}      

\begin{abstract} The perturbative analysis of models of open and closed
superstrings presents a number of surprises.  For instance, variable
numbers of antisymmetric tensors ensure their consistency via
generalized Green-Schwarz cancellations and a novel type of
singularity occurs in their moduli spaces.  All these features are
related, in one way or another, to the presence of boundaries on the
world sheet or, equivalently, of extended objects (branes) interacting
with the bulk theory in space time.  String dualities have largely
widened the interest in these models, that exhibit a wealth of generic
non-perturbative features of String Theory.
\end{abstract}

\maketitle

\section{INTRODUCTION}

The basic rules determining the perturbative spectra of oriented
closed strings essentially amount to two basic  constraints on their
torus amplitudes, {\it modular invariance} and {\it spin-statistics}
\cite{closed}. The former makes the choice of world-sheet ``time''
immaterial, while the latter fixes the relative contributions of Bose
and Fermi fields. By now familiar for almost a decade, these rules
determine the $GSO$ projections 
\cite{gso} for vacua in several space-time dimensions with different
numbers of space-time supersymmetries, and have deepened considerably
our understanding of String Theory. On the other hand, models of open
and closed strings, the very ones that marked the birth of the field
\cite{veneziano}, have long been of interest only to a small community
\cite{openold}. To some extent, this uneven fate is to be traced to
the restrictions on their gauge groups
\cite{cp} and to the relative simplicity and elegance of the heterotic
string
\cite{ghmr}.

In 1987 I suggested that a link between open-string models and {\it
symmetric} models of oriented closed strings, exhibited by the only
two examples known at the time, the bosonic string and the $SO(32)$
type-I model
\cite{gs}, be taken as a general building principle for their  spectra
\cite{cargese}. The resulting connection, a parameter-space analogue
of the orbifold construction \cite{orb}, relates the type-IIB model of
oriented closed strings
\cite{typeIIB} to its ``open descendant'', the
$SO(32)$ model of Green and Schwarz. Subsequent work exhibited the
role of boundary conditions in lower-dimensional models \cite{ps} and
led to new classes of open-string spectra in ten and six dimensions
\cite{bs}, with corresponding patterns of Chan-Paton symmetry breaking.

 Six-dimensional models promptly displayed a rewarding surprise: their
low-energy spectra include variable numbers of antisymmetric tensors
\cite{bs} that take part in a generalized anomaly cancellation
mechanism \cite{tensor}.  These marked differences with respect to
perturbative vacua of the heterotic string were correctly traced to
the ``parent'' type-IIB string. Moreover, an analysis of the
low-energy supergravity revealed the existence of peculiar
singularities in the moduli spaces of tensor multiplets, that result
in infinite gauge coupling constants \cite{tensor}.    This {\it
perturbative} phenomenon, and its  non-perturbative counterpart
expected to present itself in heterotic models
\cite{dmw}, have aroused some interest during the past year.  The
singularities reflect the presence in the vacuum of string excitations
with vanishing tension
\cite{tensionless}, indicating some sort of string generalization of
the  Nambu-Jona Lasinio-Goldstone phenomenon.  More general types of
compactifications, with non-trivial vacuum values for scalar fields in
the internal space, have been advocated to provide a geometrical
framework for these results \cite{ftheory}.

String dualities \cite{dualities}, and in particular the strong-weak
coupling relation in ten dimensions between the type-I $SO(32)$ model
and the
$Spin(32)/Z_2$ heterotic string \cite{witdual}, have turned open
strings into a central ingredient of the emerging picture. Most
notably,  Polchinski's observation
\cite{pol} that extended objects, $D$ branes, where open-string
endpoints terminate, have a central role as carriers of Ramond-Ramond
charges, has provided a conformal field theory setting for some of the
soliton studies 
\cite{dkl} that have played a central role in the development of the
string duality picture.  The recent surge of interest in open-string
vacua, accompanied by the emergence of a new term, ``orientifold'',
that stresses the geometrical significance of this peculiar orbifold
construction, has led to the construction of many more
six-dimensional  models \cite{gp,noop,tensor96,gepner}. 
Four-dimensional vacua with 
$N=1$ supersymmetry are comparatively less understood
\cite{bl,chiral}, but 
\cite{chiral} contains the first instance of a chiral model, with three
generations of chiral fermions.  The K\"ahler manifold of the
untwisted scalars in the unoriented closed sector,
$Sp(8)/(SU(4) \times U(1))$, is strongly suggestive of a
twelve-dimensional  interpretation in the spirit of \cite{ftheory}.

 The relation between closed models and their open descendants, a
general property of Conformal  Field Theory and current algebra,
associates one or more classes of ``descendants'' \cite{wzwus1,wzwus2}
to the $ADE$ series \cite{ciz} of minimal and
$SU(2)$ WZW models \cite{wzw}. The resulting models, also of some
potential relevance to Condensed Matter Physics \cite{aff}, reveal the
occurrence of another exotic phenomenon: the open sector may possess a
{\it larger} symmetry than the bulk theory \cite{wzwus2}.

This article is based on seminars presented at CERN, at the XXVIII
Institute d'Ete of the Ecole Normale Superieure in Paris and at the
XXX Ahrenshoop Symposium in Buckow. Since much literature is now
available on $D$ branes and their applications \cite{dbranes}, I shall
confine my attention to some of the ``miracles'' of open-string
perturbation theory that have shaped my own understanding of the
subject.  In the next Section I review briefly the scope of the
original proposal, while displaying, in a couple of simple instances,
the resulting structure of open-string spectra.  The following Section
illustrates the nature of the generalized Green-Schwarz mechanism,
while displaying some corresponding properties of six-dimensional
supergravity.  The final Section describes, in the open descendants of
the $D_5$ model of \cite{ciz}, the extension of internal symmetry in
the  open sector via simple currents of non-integer dimension
\cite{wzwus2}.

\section{OPEN DESCENDANTS, OR ORIENTIFOLDS}

\vskip 10pt
\noindent\input epsf \centerline{ \epsfbox{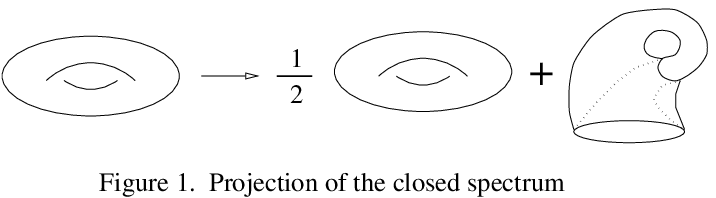}}
\vskip 10pt

Let me begin by reviewing the original proposal of \cite{cargese}.  
The basic idea (fig. 1) is to use world-sheet parity to project a
consistent spectrum of oriented closed strings symmetric under the
interchange of left and right modes.  This operation mixes left and
right modes, and its proper completion requires the Klein-bottle
amplitude.  As this violates modular invariance, singular
(ultraviolet) contributions can result from the origin, the edge of
moduli space, that in this case is a half-line.  On the other hand,
the Klein-bottle amplitude is ``dual'' to an amplitude describing the
propagation of closed strings between two cross-caps, and the modular
transformation relating these two pictures turns the ultraviolet
divergences into  infrared divergences related to massless exchanges.
A suitable spectrum of open strings, the `twisted sector'' of this
construction, can dispose of the resulting singularities, thus
restoring the consistency of the models.   Indeed, Klein-bottle,
annulus and M\"obius strip, though vastly different, are all related
by suitable modular transformations to ``dual'' amplitudes describing
closed-string exchanges between pairs of boundaries and/or
crosscaps\footnote{In order to add a crosscap to a surface, one cuts
from it a disk and glues to the resulting hole the single boundary of
a M\"obius strip.}.  In particular (fig. 2), the annulus corresponds
to a tube terminating at two boundaries, while the M\"obius strip
corresponds to a tube terminating at one boundary and one crosscap. As
a result, the proper factorization of these amplitudes turns the
consistency conditions into a set of linear equations (tadpole
conditions) for the numbers of ``colors'', or types of Chan-Paton
charges
\cite{cp}, associated to the various types of boundaries.  Some of the
tadpole conditions are related to unphysical exchanges and signal the
presence of anomalies \cite{pc}.
\vskip 10pt
\noindent
\input epsf \centerline{ \epsfbox{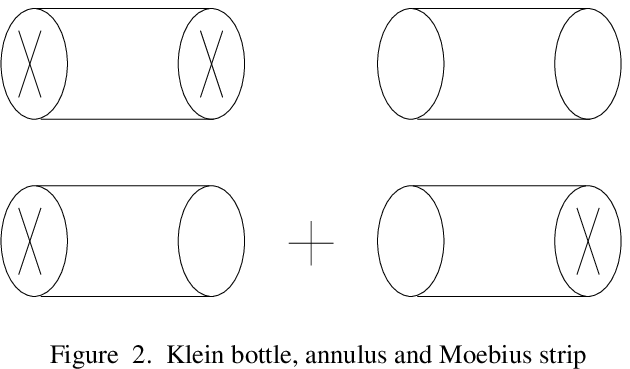}}
\vskip 10pt

It is amusing to see how this pattern is realized, leading to whole
new classes of conformal models.  Let me begin with a few simple
cases, and in particular with two classes  of non-supersymmetric
ten-dimensional models drawn from
\cite{bs}  and \cite{susy}. The models in the first class are
described by
\ba {\cal T} &=& {|O_8|}^2 + {|V_8|}^2 + {|S_8|}^2 + {|C_8|}^2 \\
{\cal K} &=& {1
\over 2} \ \left( O_8 + V_8 - S_8 - C_8 \right) \\ {\cal A} &=&
{{n_o^2 + n_v^2 + n_s^2 + n_c^2} \over 2} V_8 +  (n_o n_v + n_s n_c )
O_8  \nonumber \\ &-&(n_v n_s + n_o n_c ) S_8 - (n_v n_c + n_o n_s )
C_8 \label{ann1} \\ {\cal M} &=& - {{n_o + n_v + n_s + n_c} \over 2}
\hat{V}_8  \quad ,
\label{mob1}
\ea where the characters of level-one $SO(8)$ modules are (orthogonal)
combinations of fourth powers of Jacobi theta functions and of the
Dedekind
$\eta$ function
\ba & &{O_8 = {{\theta_3^4 + \theta_4^4} \over {2 \eta^4}}}\qquad
\qquad {V_8 =  {{\theta_3^4 - \theta_4^4} \over {2 \eta^4}}} \quad
,\nonumber
\\ & &{S_8 = {{\theta_2^4 + \theta_1^4} \over {2 \eta^4}}} \qquad
\qquad {C_8 = {{\theta_2^4 - \theta_1^4} \over {2 \eta^4}}} \quad ,
\ea and where the ``hatted'' functions, of argument $-\sqrt{q}$, are
redefined by phases and start with positive coefficients
\cite{bs,wzwus1}. In the unprojected closed sector, $|O_8|^2$ starts
with a tachyon $T$ and $|V_8|^2$ starts with the usual collection of
$g_{\mu \nu}$,
$B_{\mu \nu}$ and $\phi$.  In addition, there are two low-lying
sectors of $RR$ bosons,
${A^{+}}_{\mu \nu \rho \sigma}$, a four-form with self-dual field
strength, a two-form ${B^{\prime}}_{\mu \nu}$ and a scalar 
$\phi^{\prime}$ from $|S_8|^2$, and
${A^{-}}_{\mu \nu \rho \sigma}$, a four-form with antiself-dual field
strength, a two-form ${B^{\prime \prime}}_{\mu \nu}$ and a scalar
$\phi^{\prime\prime}$ from
$|C_8|^2$. After the projection, one is left with $g$, $\phi$ and $T$
from the NS-NS sectors, and $B^{\prime}$ and $B^{\prime \prime}$ from
the
$R-R$ sectors.  The open sector of eqs. (\ref{ann1}) and (\ref{mob1})
is chiral, since the left-handed fermions in $S_8$ and 
$C_8$ carry different types of quantum numbers.  The tadpole
conditions identify the dimensionalities of pairs of Chan-Paton charge
spaces, setting $n_o=n_v$ and
$n_s=n_c$.  The model includes {\it two} two-forms,  both involved in
the anomaly cancellation mechanism, since the residual anomaly
polynomial is
\ba I_A &\sim& \left( {\rm tr} F_o^2 - {\rm tr} F_v^2 \right) \biggl(
{\rm tr} F_s^4 - {\rm tr} F_c^4 \nonumber \\ &-&  {1 \over 4} {\rm tr}
R^2 ( {\rm tr} F_s^2 - {\rm tr} F_c^2 ) \biggr) \nonumber \\ &+&
\left( {\rm tr} F_s^2 - {\rm tr} F_c^2
\right) \biggl( {\rm tr} F_o^4 - {\rm tr} F_v^4 \nonumber \\ &-&  {1
\over 4} {\rm tr} R^2 ( {\rm tr} F_o^2 - {\rm tr} F_v^2 ) \biggr) \ .
\ea Thus, $B^{\prime}$ and $B^{\prime \prime}$ share in a democratic
fashion the crucial task of canceling the gauge anomalies, while there
are no gravitational anomalies since there is no net number of  chiral
fermions.

A variant, corresponding to a different choice of Klein-bottle
projection, was introduced in \cite{susy}.  It is particularly
interesting since it includes a
$U(32)$ model that is not supersymmetric but is free of tachyons, an
open analogue of the $SO(16) \times SO(16)$ model of \cite{so16}.  
Indeed, the choice of Klein-bottle projection, determined by the
``crosscap constraint''
\cite{wzwus1}, is {\it not} unique in general, and the different
choices correspond to vastly different spectra.  In this case, taking
\be {\cal K}={1 \over 2} \ \left( - O_8 + V_8 + S_8 - C_8 \right)
\quad ,
\ee the projected closed spectrum contains $g$ and $\phi$ (but {\it
no} tachyon) from the $NS-NS$ sectors, and $A^{+}$, $\phi^{\prime}$ and
$B^{\prime \prime}$ from the $R-R$ sectors. 

The corresponding chiral open sector, described by
\be {\cal A}\ = \ n \ \bar{n} \ V_8 \ - \ {{n^2 + {\bar{n}}^2} \over
2} \ C_8 
\label{ann32}
\ee and
\be {\cal M}= {{ n + \bar{n}} \over 2}  \ \hat{C}_8 \quad ,
\ee can be obtained permuting some of the charge assignments in eq.
(\ref{ann1}) and turning pairs of ``real'' charge spaces into pairs of
``complex'' spaces for the (anti)fundamental representations of a
unitary group.  The corresponding multiplicities are equal in pairs,
and in this case $n = \bar{n} =32$. This recurrent possibility,
related to the presence of simple  currents\footnote{Simple currents
are operators with abelian fusion rules with all other fields in the
theory.  This notion was introduced in \cite{sy}.} in the spectrum, 
was first noted in \cite{bs} and is studied in detail in
\cite{wzwus1}.

The residual anomaly polynomial of this model
\ba I_A &\sim& \left( {\rm tr} F^2 - {\rm tr} R^2 \right) \biggl( {\rm
tr} F^4 \ +
\ {1 \over 8} \ {\rm tr} R^4 \nonumber \\ &+& {1 \over 32} \ {( {\rm
tr} R^2 )}^2
\ - \ {1 \over 8} \ {\rm tr} F^2 \ {\rm tr} R^2 \biggr) \nonumber \\
&+& {2 \over 3} {\left( {\rm tr} F^3 \ - \ {1 \over 8} \ {\rm tr} F \
{\rm tr} R^2 \right)}^2
\nonumber \\ &+& {\rm tr} F \biggl( {2 \over 5} \ {\rm tr} F^5 \ + \
{1 \over {240}} {\rm tr} F \ {\rm tr} R^4 \nonumber \\ &-& {1 \over
{192}} \ {\rm tr} F \ {( {\rm tr} R^2 )}^2 \biggr)
\label{apoly2}
\ea is particularly interesting.  Together with the two-form $B^{\prime
\prime}$, two more fields take part in a generalized Green-Schwarz
mechanism,
$A^{+}$ and the eight-form dual to $\phi^{\prime}$.  The coupling of
the latter gives mass to the $U(1)$ vector field, and the actual
low-energy gauge group is thus $SU(32)$.  It should be noticed, in
particular, that the self-duality of the field strength of
$A^{+}$ reflects itself in the nature of the corresponding term in eq.
(\ref{apoly2}), a perfect square.  We have thus come across the first
miracle alluded to in the Introduction: the Green-Schwarz mechanism
takes typically a more general form in these models, with
contributions from forms of different degrees.  This is actually the
most complicated example I am aware of, since the six-dimensional ones
of \cite{tensor} only include (anti)self-dual two-forms, and
occasionally four-forms, in a way that extends the original
four-dimensional mechanism of
\cite{witt4}.

Let me now turn to a couple of additional instances.  The first
presents itself when one compactifies the type-I superstring to nine
dimensions, and will serve the purpose of illustrating the role of $D$
branes in connection with
$T$-duality. In this case
\be {\cal T} \ = \ {| V_8 - S_8 |}^2 \ {{\sum \ q^{{\alpha^{\prime}
\over 4} p_L^2}
\ \bar{q}^{{\alpha^{\prime} \over 4} p_R^2}} \over {\eta(q)
\bar{\eta}(\bar{q})}}
\quad ,
\label{ttor}
\ee where, as usual,
\be p_{L,R} \ = \ {m \over R} \ \pm \ {{n R} \over {\alpha^{\prime}}}
\quad ,
\ee describes the toroidal compactification of the Type-IIB
superstring to nine dimensions. In the absence of Wilson lines, the
familiar Klein-bottle projection
\be {\cal K}= {1 \over 2} \ \left( V_8 - S_8 \right) \ {{\sum \ 
q^{{\alpha^{\prime}} {{m^2} \over {2 R^2}}} } \over {\eta( q^2 )}} 
\label{ktor}
\ee leads to
\be {\cal A} = {N^2 \over 2} \ \left( V_8 - S_8 \right) \ {{\sum \ 
q^{{\alpha^{\prime}} {{m^2} \over {2 R^2}}} } \over {\eta(\sqrt{q})}} 
\label{ator}
\ee and
\be {\cal M}= \ - \ {N \over 2} \ \left( \hat{V}_8 - \hat{S}_8 \right)
\  {{\sum
\ q^{{\alpha^{\prime}} {{m^2} \over {2 R^2}}} } \over {\hat{\eta}( -
\sqrt{q} )}}
\quad ,
\label{mtor}
\ee where $N=32$ and, again, the power series of ``hatted'' functions
starts with unit coefficient. The spectrum is clearly {\it not}
invariant under $T$ duality, since ${\cal K}$, ${\cal A}$ and ${\cal
M}$ are all drastically altered by the usual replacement of $R$ with
$\alpha^{\prime}/ R$.  This should be contrasted with the familiar
case of the heterotic string, where $T$ duality finds an elegant
setting in the Narain lattice construction \cite{nsw}. One  can
actually explain in rather simple terms what is  happening
\cite{tdual}.  Indeed, $T$ duality is an asymmetric parity operation
on the world sheet, that transforms  the (anti)holomorphic portions of
the internal string coordinate according to
$X_L
\rightarrow X_L$ and $X_R \rightarrow - X_R$.  Since the usual Neumann
condition essentially associates the open-string sector to
$X_L + X_R$, a $T$ duality turns it into a "Dirichlet" open sector
corresponding to $X_L - X_R$.

Where do the ``Dirichlet'' ends terminate? We owe to  \cite{pol} a
pervasive geometrical characterization of these extended objects, now
commonly referred to as $D$ branes.  In this  case, they are just
hyperplanes, but a closer inspection reveals that they are actually
dynamical objects, a variety of string solitons.  There is already a
vast literature on this important topic
\cite{dbranes}, and I shall thus confine my attention, as advertised,
to their role in {\it perturbative} open-string spectra.  Indeed, the
first lesson is precisely that they {\it are} part of perturbative
open-string spectra, so what better way do we have to gain some
familiarity with their properties!  The second observation  is that,
allowing for these objects, one is apparently going past the original
restriction \cite{cargese} to models with a left-right symmetric GSO
projection.  Actually, insofar as one is  ready to perform suitable
$T$-duality transformations, every situation is captured by the
original setting. There is a simple, canonical example, that I find
quite instructive. It is a generalization of the toroidal model of
eqs. (\ref{ator})-(\ref{mtor}) where, following
\cite{torus}, one allows for Wilson lines breaking $SO(32)$ to
$U(M) \times SO(32 - 2 M)$.  The Wilson lines affect the energy levels
via minimal couplings, so that
\ba {\cal A} &=& \left( {{N^2} \over 2} + M \bar{M} \right) \ 
\left( V_8 - S_8 \right)  {{\sum \ q^{{\alpha^{\prime}} {{m^2} \over
{2 R^2}}} }
\over {\eta(\sqrt{q})}} \nonumber \\ &+& N M \ \left( V_8 - S_8
\right) \ {{\sum
\ q^{{\alpha^{\prime}}  {{(m + a R/2)^2} \over {2 R^2}}} } \over
{\eta(\sqrt{q})}}
\nonumber \\ &+& N \bar{M} \ \left( V_8 - S_8 \right) \ {{\sum \
q^{{\alpha^{\prime}}  {{(m - a R/2)^2} \over {2 R^2}}} } \over
{\eta(\sqrt{q})}}
\nonumber \\ &+& {M^2 \over 2} \ \left( V_8 - S_8 \right) \ {{\sum \
q^{{\alpha^{\prime}}  {{(m + a R)^2} \over {2 R^2}}} } \over
{\eta(\sqrt{q})}}
\nonumber \\ &+& {\bar{M}^2 \over 2} \ \left( V_8 - S_8 \right) \
{{\sum \ q^{{\alpha^{\prime}}  {{(m - a R)^2} \over {2 R^2}}} } \over
{\eta(\sqrt{q})}}
\quad ,
\label{awilson}
\ea
\ba {\cal M}&=& - \ {N \over 2} \ \left( \hat{V}_8 - \hat{S}_8 \right)
\  {{\sum
\ q^{{\alpha^{\prime}} {{m^2} \over {2 R^2}}} } \over {\hat{\eta}( -
\sqrt{q} )}}
\nonumber \\
 &-& {M \over 2} \ \left( \hat{V}_8 - \hat{S}_8 \right) \  {{\sum \
q^{{\alpha^{\prime}} {{{(m + a R )}^2} \over {2 R^2}}} } \over
{\hat{\eta}( -
\sqrt{q} )}} \nonumber \\
 &-& {\bar{M} \over 2} \ \left( \hat{V}_8 - \hat{S}_8 \right) \ 
{{\sum \ q^{{\alpha^{\prime}} {{{(m - a R )}^2} \over {2 R^2}}} }
\over {\hat{\eta}( -
\sqrt{q} )}} \quad .
\label{mwilson}
\ea where  $N + 2 M = 32$.

The M\"obius contribution is only associated to terms of ${\cal A}$
with identical charges, but the corresponding  momentum shifts are
doubled, since in the proper  parametrization its boundary has twice
the length of each boundary of the annulus. All this is nicely
consistent with the factorization of ``dual'' vacuum channels.  

If one performs a
$T$ duality transformation, the translations in momentum space turn
into translations in coordinate space.   As these are simple to
visualize (fig. 3), the resulting $D$-brane picture, though
equivalent, has undoubtedly some aesthetic appeal.   For instance, it
is evident from eqs. (\ref{awilson}) and (\ref{mwilson}) that
$a = {1 \over R}$ results in an enhancement (or rather, a partial
recovery) of the gauge symmetry, that becomes $SO(32-2M) \times
SO(2M)$, while $a={2
\over R}$ retrieves the full $SO(32)$ gauge group.  Referring to fig.
3, in the
$T$-dual picture the prototype breaking of eqs. (\ref{awilson}) and
(\ref{mwilson}) involves strings stretched between a $D$ brane $(D_o)$
at  the bottom of the circle and an additional pair of $D$ branes at
points symmetrically located with respect to the vertical
$( D_{+}$ and $D_{-} )$.  The partial recovery of gauge symmetry takes
place when the last two meet at the top of the circle, whereas full
recovery is attained when all three meet at the bottom.
\vskip 8pt
\noindent\input epsf \centerline{ \epsfbox{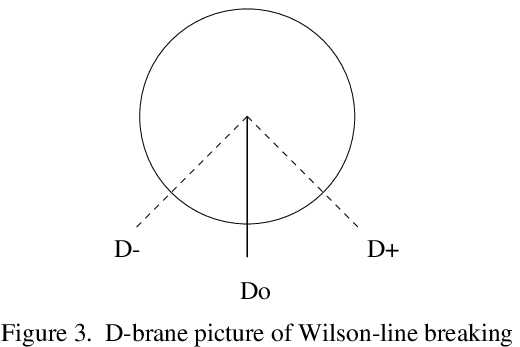}}
\vskip 10pt

One can actually modify the Klein-bottle projection, since the crosscap
constraint
\cite{wzwus1} allows one to (anti)symmetrize internal momenta
according to their quantum numbers. This interesting possibility,
first exploited in 
\cite{noop} and discussed further in \cite{gepner}, corresponds to
\be{\cal K}= {1 \over 2} \ \left( V_8 - S_8 \right) \ {{\sum \ {(-1)}^m
q^{{\alpha^{\prime}} {{m^2} \over {2 R^2}}} } \over {\eta( q^2 )}} 
\label{kexh}
\ee and leads to open descendants {\it without} an open sector.
Indeed, familiar properties of theta functions imply that the
transverse-channel amplitude corresponding to eq. (\ref{kexh}) does
not introduce any singularities, since it does not  involve any
massless exchange.

Let me now complicate matters slightly.  Following \cite{ps}, I shall
describe how to associate open descendants to $c=1$ orbifolds. While
the resulting conformal models can be associated to string
compactifications, for simplicity I shall only consider the internal
part. The novelty is that strings with Neumann and Dirichlet
conditions coexist in the spectrum, independently of
$T$ dualities.  This geometrical setting has been discussed for the
superstring in
\cite{gp}, with resulting constraints equivalent to those in
\cite{ps}. The covariant form has the advantage of leading naturally
to large Chan-Paton gauge groups.  Indeed, in rational models one
typically faces the presence of {\it quantized} backgrounds for the
$NS-NS$ two-form, and these reduce the rank of the Chan-Paton group. 
For instance, starting from a total dimensionality $32$ for the
Chan-Paton charge space, a quantized two-form of (even) rank $r$
results in effective quantized Wilson lines that reduce the
dimensionality of the charge space to $32/2^{r/2}$
  \cite{torus}.

The oriented closed sector \cite{orb} is built projecting the internal
part of the torus amplitude (\ref{ttor}) so as to enforce the
identification  between
$X$ and
$-X$ required by the geometry of the orbifold.  The resulting
``untwisted'' sector breaks modular invariance, that is recovered
associating a pair of ``twisted'' sectors to the two fixed points of
the orbifold.  The resulting torus amplitude is
\ba {\cal T} &=& {1 \over 2} \ {{\sum \ q^{{\alpha^{\prime} \over 4}
p_L^2}
\ \bar{q}^{{\alpha^{\prime} \over 4} p_R^2}} \over {\eta \bar{\eta}}} 
+ {1 \over 2} \left| {{\theta_3 \theta_4} \over {\eta}^2} \right|
 \nonumber \\ &+& {1 \over 2} \left| {{\theta_2 \theta_3} \over
{\eta}^2}
\right| + {1 \over 2} \left| {{\theta_2 \theta_4} \over {\eta}^2}
\right| \quad .
\label{torb}
\ea

In order to construct the unoriented closed sector of the descendant, 
one starts again from the Klein bottle amplitude.  In this case,
however, an inspection of the sections allowed on the genus-one
surfaces \cite{ps} or, equivalently, of the operator content, reveals
the need for both momentum and winding sums. As a result 
\ba {\cal K} &=& {1 \over 4} \  {{\sum_m \ q^{\alpha^{\prime} {m^2
\over {2R^2}}}}
\over {\eta(q^2)}} 
\ + \ {1 \over 4} \ {{\sum_n \ q^{{n^2 R^2} \over {2
\alpha^{\prime}}}} 
\over {\eta(q^2)}} \nonumber \\ &+& {\left({{\theta_2 \theta_3} \over
{\eta}^2} ( q^2 ) \right)}^{1/2}
\label{korb}
\ea is invariant under $T$ duality transformations.  The unitarity of
the vacuum channels exposed by an $S$ transformation then requires
that the open sector contain momentum and winding sums as well. 
Therefore, one has to face the simultaneous presence of $NN$ and $DD$
strings, since the former can have momentum, while the latter can have
windings in the internal direction.   It should be appreciated,
however, that the endpoints of these $DD$ strings sit at the two fixed
points of the orbifold.  There are thus three types of $DD$ strings:
with both ends at one or the other fixed point, or stretched between
the two fixed points.  This is a discrete version of the breaking of
fig. 3, as pertains to the absence of corresponding moduli in the
projected spectrum. Aside from the usual
$NN$ strings (possibly with ``quantized'' Wilson lines), there are
additional ones with one $N$ and one $D$ end, with the latter at any
of the two fixed points.  In conclusion,
\ba {\cal A} &=& {{tr({\bf 1}_N )}^2 \over 4} \ {{\sum_m \ 
q^{\alpha^{\prime}{m^2
\over {2R^2}}}} \over {\eta}} \nonumber \\ &+& {{tr({\bf R}_N )}^2
\over 4} \ {\left({{\theta_3 \theta_4} \over {\eta}^2} 
 \right)}^{1/2} \nonumber \\ &+& 
\sum_{i=1,2} \ {{tr({\bf 1}_N ) tr({\bf 1}_{Di} )} \over 4} \
{\left({{\theta_2
\theta_3} \over {\eta}^2} 
 \right)}^{1/2} \nonumber \\ &+&
\sum_{i=1,2} \ {{tr({\bf R}_N ) tr({\bf R}_{Di} )} \over 4} \
{\left({{\theta_2
\theta_4} \over {\eta}^2} 
 \right)}^{1/2} \nonumber \\ &+&
\sum_{i,j=1,2} \ {{tr({\bf 1}_{Di} )}^2 \over 4} \ {{\sum_n \ 
q^{\alpha^{\prime}{{(n + (i-j)/2)R}^2 \over {2 \alpha^{\prime}}}}} 
\over {\eta}} \nonumber \\ &+&
\sum_{i=1,2} \ {{tr({\bf R}_{Di} )}^2 \over 4} \ {\left({{\theta_3
\theta_4}
\over {\eta}^2} 
 \right)}^{1/2}
\label{aorb}
\ea and
\ba {\cal M}&=& {{tr( {\bf 1}_N ) + \sum_{i=1,2} \ tr( {\bf 1}_{Di} )}
\over 4} {\left({{\hat{\theta}_3 \hat{\theta}_4} \over {\hat{\eta}}^2} 
 \right)}^{1/2}
\nonumber \\ &+& \sum_{i=1,2} \ {{tr( {\bf 1}_{Di} )} \over 4} \
{{\sum_n \  q^{\alpha^{\prime}{{(n + (i-j)/2)R}^2 \over {2
\alpha^{\prime}}}}} 
\over {\hat{\eta}}} \nonumber \\ &+&  {{tr( {\bf 1}_N )} \over 4} \
{{\sum_m \  q^{\alpha^{\prime}{m^2 \over {2R^2}}}} \over {\hat{\eta}}}
\ ,
\label{morb}
\ea  where $tr( {\bf 1} ) = Q_{+} + Q_{-}$ and $tr( {\bf R} ) = Q_{+}
- Q_{-}$ implement a $Z_2$ breaking in the charge sectors.  For
brevity, I have omitted the arguments of all $\theta$ and $\eta$
functions, equal to
$\sqrt{q}$ in the annulus amplitude and to $-\sqrt{q}$ in the M\"obius
amplitude.

It can be verified that the vacuum channels  of ${\cal K}$ and ${\cal
A}$ comprise contributions of the  different sectors with squared
reflection coefficients, and that, as usual
\cite{bs}, the coefficients in vacuum channel of ${\cal M}$ are
geometric  means of them.  All this, of course, if the $D$ charges sit
at the fixed points, and if a corresponding quantization holds for the
Wilson lines acting on $N$ charges, that I have not displayed in eq.
(\ref{morb}). In space-time models,  tadpole conditions related to
untwisted massless exchanges constrain the dimensionality of the
charge space, whereas others related to twisted exchanges constrain
the breaking pattern.  Actually, the choice in eqs. (\ref{aorb}) and
(\ref{morb}) is not unique, and an additional one may be obtained
interchanging $Tr( {\bf 1} )$ and $-Tr( {\bf R} )$, while identifying
pairs  of Chan-Paton charge spaces.  This is one more instance of the
phenomenon discussed at length in \cite{bs} and
\cite{wzwus1} and already encountered in eq. (\ref{ann32}), whereby
pairs of ``real'' charge spaces associated to orthogonal or symplectic
groups are traded for pairs of conjugate spaces of ``complex'' charges
associated to unitary groups.  

In the open sector $T$ duality is not necessarily a symmetry, since
it  involves the interchange of 
$N$ and $D$ charges, but it can be if identical breaking patterns are
chosen for the $N$ and $D$ charge spaces.   In the former, the
breaking is controlled by (quantized) Wilson lines, while in the
latter it is determined by the distribution of $D$ charges among the
two fixed points. This has been discussed at length in
\cite{gp,noop,tensor96}. In modern parlance, this model exhibits the 
coexistence, again in the {\it perturbative} spectrum, of $D$ branes of
different dimensions.  Part of this structure was anticipated long ago
in the low-energy analysis of \cite{hm}, that motivated the string
analysis of
\cite{ps}.

\section{TENSOR MULTIPLETS IN SIX DIMENSIONS}

Let me now turn to six dimensions, and in particular to type-I models
with  simple supersymmetry.  They are descendants of the {\it unique}
$K3$ reduction of the Type-IIB string \cite{k3}, whose low-energy
dynamics is described by
$N=2b$ supergravity \cite{romans} with 21
$N=2b$ tensor multiplets\footnote{In this article I am abiding to
present-day notation, whereas in
\cite{romans} and \cite{tensor} simple supersymmetry was referred to
as $N=2b$.}  and a moduli space $SO(5,21)/SO(5) \times SO(21)$. There
are only two relevant
$N=2b$ multiplets: the gravitational multiplet includes two
left-handed Weyl gravitini, the gravitational field and five self-dual
two-tensors, while the tensor multiplet includes a single
antiself-dual tensor, two right-handed Weyl spinors and five scalar
fields. The
$N=2b$ supergravity was constructed, to lowest order in Fermi fields,
in
\cite{romans}, while the unique anomaly-free spectrum was identified in
\cite{agw}. 

In proceeding to the open descendants, the novelty is that the
projection can result in models with simple supersymmetry in six
dimensions with  different numbers of residual two-tensors.  In simple
superymmetry, the gravitational multiplet includes the gravity field,
a left-handed Weyl gravitino and a self-dual two-form, while the
tensor multiplet includes an antiself-dual two-form, a right-handed
Weyl spinor and a scalar.  In this article we shall also consider the
vector multiplet, that includes a  vector and a left-handed Weyl
spinor.  

There is a particularly simple class of models, originally constructed
in
\cite{bs}, where I first came across the generalized Green-Schwarz
mechanism.  Many similar models have been recently discussed in
\cite{noop,tensor96,gepner}. In particular, \cite{gepner} discusses
the only  string vacuum I am aware of with {\it no} tensor multiplets
at all.  Our simple example is built starting from the toroidal
reduction of the Type-IIB string on the
$SO(8)$ lattice. In the notation of the  previous Section, the torus
amplitude reads
\be {\cal T} = {| V_8 - S_8 |}^2 \left( {| O_8 |}^2 + {| V_8 |}^2 + {|
S_8 |}^2 +  {| C_8 |}^2 \right).
\ee The next step is a conventional $Z_2$ orbifold \cite{orb}, where
one reverses simultaneously the sign of the four internal fermions
$\psi^i$ and  of four of the eight internal fermions that bosonize the
internal coordinates.  Precisely in the spirit of the fermionic
construction in
\cite{closed}, this is manifestly compatible with the corresponding
fermionic supercurrent.  The characters break accordingly into pairs
with opposite eigenvalues under the $Z_2$ projection.  Thus, for
instance,
\be V_8 - S_8 \rightarrow ( V_4 0_4 - C_4 C_4 ) + ( 0_4 V_4 - S_4 S_4
) \ ,
\ee where the two terms, denoted $Q_o$ and $Q_v$ in \cite{bs} have,
respectively,  positive and negative eigenvalues under the involution,
and the resulting torus amplitude for the Type-IIB string on the
orbifold,
\be {\cal T} \ = \ \sum_{i=1}^{16} \ {| \chi_i |}^2 \quad ,
\label{tor16}
\ee
 is a diagonal combination of 16 characters.  The identity is
\be
\chi_1 \ = \ Q_o \ O_4 \ O_4 \ + \ Q_v \ V_4 \ V_4 \quad ,
\ee and the remaining characters are simply determined after a modular
$S$ transformation of this expression.  Building the ``standard''
descendant is also simple, since the ``parent'' model is diagonal. 
Indeed, in this case Cardy
\cite{cardy} has related the spectrum of boundary operators to the
fusion-rule coefficients, an observation turned in \cite{bs} into a
set of rules for building descendants of rational models. Thus, the
open sector corresponding to (\ref{tor16}) includes 16 charge sectors
and
\be {\cal A} \ = \ {1 \over 2} \ \sum_{ijk} \ n^i \ n^j \ N_{ij}^k \
\chi_k
\quad ,
\label{adiag}
\ee where $N_{ij}^k$ are the fusion-rule coefficients.  Taking into
account the  five tadpole conditions, one is left with a number of
models, the simplest of which has gauge group $Sp(8)^{\otimes 4}$.  

The closed spectrum of this class of models includes five tensor
multiplets, and a direct analysis of the anomaly polynomial led to a
big surprise. Its irreducible part cancels as usual after imposing the
tadpole conditions, but one is left with the residual contribution
\be I_A \ \sim \ \sum_{x,y} \ \eta_{rs} \ c^r_x \ c^s_y \  Tr_x F^2 \ 
Tr_y F^2
\quad ,
\ee  where the labels refer to the simple factors of the gauge group, 
that {\it does not} factorize.  Here
$\eta$ is a Minkowski metric of signature
$(1 - n_T)$, with $n_T$ the number of tensor multiplets, and the $c's$
are constants. I have not indicated explicitly the gravitational
contribution, but this could be done extending the range of $(x,y)$,
at the price of including corresponding higher-derivative terms.  This
setting is thus  more general than the original one considered by
Green and Schwarz
\cite{gs}, since it demands Chern-Simons couplings for {\it several}
antisymmetric tensors.  Moreover, since the anomaly polynomial is
quadratic,  for the gauge part the mechanism is realized at the level
of two-derivative terms.  The analysis of the corresponding low-energy
supergravity is thus bound to  present some peculiarities, and indeed
it does.  In the first of
\cite{tensor} I constructed its field equations to lowest order in the
Fermi fields requiring that the resulting effective action be
invariant under supersymmetry. On the other hand, in the second paper
of \cite{tensor} the field equations are derived from the Wess-Zumino
consistency conditions \cite{wz}, extending previous studies of
anomalous models in global supersymmetry
\cite{anomglob}. The choice affects the gauge current, whose
divergence gives in the former case the ``covariant'' anomaly and in
the latter the more satisfactory ``consistent'' anomaly.  The current
is {\it not} conserved, since in this model fermion loops {\it are to}
cancel the residual, factorized anomaly.  Of course, this is just
Noether's theorem at work, and similar considerations apply to the
Green-Schwarz setting of
\cite{gs}, where this inconsistency is present in higher-derivative
couplings.  

For each choice for the gauge current $J$, the vector field equation
takes the form
\be v_r \ c^r_z \ D_{\mu } \ {F_z}^{\mu \nu} \ = \ {J_z}^{\nu}
\label{vector}
\ee for each factor $z$ of the gauge group, where $v_r$ are (redundant)
coordinates for the scalar fields of the tensor multiplets, that live
on the hyperboloid
\be
\eta^{rs} \ v_r \ v_s \ = \ 1 \quad .
\label{hyperb}
\ee Since this describes the coset $SO(1,n_T )/SO( n_T )$, the
$SO(1,n_T)$ symmetry is spontaneously broken in the vacuum to $SO( n_T
)$.  Even the latter symmetry, however, is affected by the couplings
$c^r_z$ induced by the anomaly, and as result the effective gauge
couplings 
\be {1 \over {g_{(z)}^2}} \ = \ v_r \  c^r_z
\ee in the vector kinetic term corresponding to (\ref{vector}),
\be L \ = \ - {e \over 2} \ v_r \ c^r_z \ tr_z \left( F_{\mu \nu} \ 
F^{\mu \nu}
\right) \quad ,
\ee  can blow up at some points of moduli space \cite{tensor}.  This
unusual feature, exploited in \cite{dmw} in the context of
six-dimensional heterotic-heterotic dualities, was ascribed in
\cite{tensionless} to a novel phenomenon: the singularities signal the
presence, in the  vacuum, of a new type of excitation, a tensionless
string. The argument is based on the structure of six-dimensional
central charges and on the second-order tensor equation of
\cite{tensor}, whose source term is the instanton density.  In this
analogue of the familiar Nambu-Jona Lasinio-Goldstone transition,
where a particle becomes massless, one expects a whole tower of
higher-spin excitations to play a role. This is the second of the
peculiarities alluded to in the Introduction.  

Let me stress once more that the vector couplings to the scalars of
tensor multiplets induced by the anomaly are revealed by a {\it
perturbative} analysis of type-I vacua, where the dilaton sits in a
hypermultiplet. On the other hand, they are quite surprising if
referred to the heterotic string, where the dilaton sits in a tensor
multiplet, since the vector kinetic terms  corresponding to eq.
(\ref{vector}) contain non-perturbative  contributions
\cite{dmw,tensionless}. For instance, if a single tensor multiplet is
present one can solve the constraint  of eq. (\ref{hyperb}) in terms
of hyperbolic functions of a scalar. In the heterotic case, this is to
be identified with the dilaton, and the vector kinetic term
corresponding to eq. (\ref{vector})
\be L_V = \ - \ {e \over 2} \ \left( c_0 e^{- \phi} + c_1 e^{\phi}
\right) {\rm tr} \ F_{\mu \nu } F^{\mu \nu}
\ee includes the non-perturbative coupling described by $c_1$.

A simple formula, first derived in \cite{chiral},  follows from the
ten-dimensional duality conjecture of \cite{witdual} and relates the
dilatons of heterotic and Type-I vacua:
\be {\phi_I}^{(d)} = {{6 - d} \over 4} {\phi_H}^{(d)} - {{d-2} \over
16} \
\log \det {G_H}^{(10-d)} \; .
\ee It should be noted that the two dilatons are independent precisely
in six  dimensions. Though quite surprising from the heterotic
viewpoint, it is thus natural to construct Type-I vacua {\it without}
tensor multiplets \cite{gepner}.  More amusingly, one is naturally led
to speculate that, in their duals,  the ``missing'' dilaton has been
effectively stabilized by non-perturbative effects.  This point is
worthy of a closer look, since it would realize in a simple setting a
phenomenon long advocated for four-dimensional string physics.

\section{THE SYMMETRY OF BOUNDARY STATES}

I shall conclude with a brief discussion of the  last exotic phenomenon
mentioned in the Introduction,  while confining my attention to the
rational descendants of a rational conformal theory. In principle, one
would expect that the open sector, determined by reflections at the
ends of the tube,  inherit at most the diagonal part of the symmetry
algebra (Virasoro or extended) $A \times
\bar{A}$ of the bulk theory.  Rather surprisingly, however, in some
instances the symmetry of boundary states turns out to be larger. 
This phenomenon was first  noted \cite{wzwus2} in the open descendants
of the $D_5$ model of the
$ADE$  classification of \cite{ciz}.

It is  convenient to assume that the  bulk spectrum has been resolved
into independent components, so that the torus partition function  is
a permutation invariant built out of $m$ characters, 
\be {\cal T} = \sum_{i=1}^m \ \bar{\chi}_i \ \chi_{\sigma(i)} \quad .
\ee The relevant techniques have been developed in \cite{fss}.  In the
diagonal case\footnote{Diagonal means, effectively, that
${\cal T}$ is built from the charge-conjugation modular invariant,
{\it i.e.}
$\sigma(j)=c(j)$.}, the canonical choice for the annulus of eq.
(\ref{adiag})
\cite{cardy,bs} reflects the one-to-one correspondence between chiral
sectors of the bulk spectrum and boundary states. It is also nicely 
consistent with factorization \cite{bs}, since the Verlinde formula
guarantees that in the corresponding vacuum channel 
$\tilde{\cal A}$, exposed by an $S$ modular transformation,  each
sector enters with a coefficient that is a perfect square:
\be {\tilde{\cal A}} \ \sim \ \sum_i \ \chi_i \ {\left( \sum_j
{{S_{ij} n^j}
\over
\sqrt{S_{1i}}} \right)}^2 \quad .
\label{atdiag}
\ee 

In Conformal Field Theory, the annulus amplitude may be regarded as a
generating function for the multiplicities of boundary operators that
mediate between boundary states. Thus, in the diagonal case each
chiral sector has
$m$ independent reflection coefficients for the $m$ choices of
boundary states at the ends of the tube, all expressible in terms of
the $S$ matrix.  

The $A_7$ model of \cite{ciz}
\ba {\cal T}_{A_7} &=& {| \chi_1 |}^2 + {| \chi_2 |}^2 +{| \chi_3 |}^2
+{|
\chi_4 |}^2 + {| \chi_5 |}^2\nonumber \\ &+& {| \chi_6 |}^2 + {|
\chi_7 |}^2
\quad ,
\label{ta7}
\ea where the labels of the seven level-six characters are $2 I + 1$,
with $I$ their isospins, is a convenient example with unextended
symmetry. On the other hand, the $D_6$ model of \cite{ciz} has level
eight and
\be {\cal T}_{D_6} = {| \chi_1 + \chi_9 |}^2 + {| \chi_3 + \chi_7 |}^2
+ 2 {|
\chi_5 |}^2 \quad ,
\label{tdeven}
\ee is a simple model with extended symmetry. In this case, there is a
fixed-point ambiguity, resolved assigning {\it two} different fields
to the term with multiplicity 2.  Together with the other combinations
of level-eight
$SU(2)$  characters in eq. (\ref{tdeven}), these define a fusion
algebra with some
$N_{ij}^k$ larger than one.  As a result, some terms occur in ${\cal
A}_{D_6}$ with multiplicities.  This result reflects the existence of
inequivalent three-point functions.  The simplest instance of this
phenomenon is quite familiar: in $SU(3)$ there are two invariant
tensors with three adjoint indices, the structure constants $f^{abc}$
and the symmetric tensor $d^{abc}$.

In open descendants of off-diagonal models, the direct correspondence
between boundary states and chiral sectors of the bulk spectrum is
lost, since in this case the tube can only support a portion of them. 
In particular,  {\it if the boundaries are to preserve the bulk
symmetry, the allowed sectors are those paired with their conjugates
in the torus modular invariant.}  Thus, some of the reflection
coefficients vanish, a condition naturally met by suitable linear
combinations of the ``diagonal'' boundary states. However, {\it not
all boundary states are obtained in this way}.  For instance, in all
$D_{odd}$ models one has a simple current of half-integer dimension,
and the proper boundary states are combinations of pairs of
``diagonal'' ones invariant under the action of this simple current
\cite{wzwus2}.  This is quite reminiscent  of the structure of the
first two terms in eq. (\ref{tdeven}), and is effectively an extension
by the simple current, that in these models has non-integer dimension!
As in the more familiar bulk case, the boundary state naively
invariant under the action of the simple current actually splits into
a pair of states.  For instance, in the
$D_5$ case the torus modular invariant is built out of the level-six
characters of eq. (\ref{ta7}), but
\ba {\cal T}_{D5} &=& {| \chi_1 |}^2 + {| \chi_3 |}^2 + {| \chi_5 |}^2
+  {|
\chi_7 |}^2 \nonumber \\ &+& {|\chi_4|}^2 + \chi_2 \bar{\chi}_6 +
\chi_6
\bar{\chi}_2 \quad ,
\ea  and therefore one would expect a total of five boundary states. 
However, there are only three combinations of ``diagonal'' boundary
states invariant under the action of the simple current, $(1+7)$,
$(2+6)$ and $(3+5)$. The correct total number, five, is thus recovered
precisely because the state corresponding to the middle field splits
into a pair $4_{+}$ and $4_{-}$.  The construction in \cite{wzwus1}
recovers quite naturally the pair of split fields, but there is also
an analogue of the multiple  three-point functions of conventional
bulk extensions.  This is directly induced by the fusion of pairs of
``off-diagonal'' boundary states, and has a direct consequence for the
annulus amplitude, that contains multiplicities. For instance, the
conventional (real) choice for the $D_5$ annulus amplitude
\ba
\lefteqn{2 A_{D_5} \ =  \
\chi_1 \bigl( n_1^2 + n_2^2 + n_3^2 + n_{4_{+}}^2 + n_{4_{-}}^2 
\bigr) \ +} \nonumber \\
\lefteqn{ (\chi_2 + \chi_6 )\bigl( 2 ( n_1 n_2  + n_2 n_3 +  n_{4_{+}}
n_3 +  n_{4_{-}} n_3)\bigr) \ +} \nonumber \\
\lefteqn{ \chi_3 \bigl( n_2^2 + 2 n_3^2 + 2( n_2 n_{4_{+}} +  n_2
n_{4_{-}} \ +}
\nonumber \\
\lefteqn{n_{4_{+}} n_{4_{-}} +   n_1 n_3)\bigr) + \chi_{4} \bigl( 4
n_2 n_3 +  2(n_1 n_{4_{+}} \ +} \nonumber \\
\lefteqn{ n_{4_{+}} n_3 + n_1 n_{4_{-}} +  n_{4_{-}} n_3 )\bigr)  + 
\chi_5 \bigl(n_2^2  + n_{4_{+}}^2 \ +}
\nonumber \\ 
\lefteqn{n_{4_{-}}^2 + 2 n_3^2 + 2(n_2  n_{4_{+}} +
 n_2 n_{4_{-}} +  n_1 n_3 )\bigr) \ +} \nonumber \\
\lefteqn{  \chi_7 \bigl( n_1^2 + n_2^2 + n_3^2 +  2 n_{4_{+}}
n_{4_{-}} \bigr)
\label{ard5} \quad , }
\ea contains multiplicities in $\chi_3$ and $\chi_5$, and the whole
$D_{odd}$ series follows a similar pattern.
\vskip 12pt

 I would like to thank the Organizers of the XXX Ahrenshoop Meeting
for their kind invitation to present this material.  I am very
grateful to Massimo Bianchi, Gianfranco Pradisi and Yassen S. Stanev
for a long and enjoyable collaboration.  I am also grateful to Carlo
Angelantonj, Sergio Ferrara, Ruben Minasian and Fabio Riccioni for
extensive discussions and an enjoyable collaboration over the past
year.

\end{document}